\documentclass[a4paper,10pt]{article}
\usepackage{graphicx}
\usepackage{epsfig}
\usepackage[american]{babel}
\usepackage{amssymb}
\usepackage{calc}

\begin{document}

\title{Bernoulli mapping with hole and a saddle-node scenario of the birth
of hyperbolic Smale--Williams attractor}
\author{Olga~B.~Isaeva$^{1,2}$, Igor~R.~Sataev$^1$}
\date{}
\maketitle\begin{center} \emph{
$^1$Kotel'nikov's Institute of Radio-Engineering and Electronics of RAS, Saratov Branch \\
Zelenaya 38, Saratov, 410019, Russian Federation}\end{center}

\maketitle\begin{center} \emph{
$^2$Saratov State University \\
Astrakhanskaya 83, Saratov, 410026, Russian Federation}\end{center}

\begin{abstract}
One-dimensional Bernoulli mapping with hole is suggested to describe the
regularities of the appearance of a chaotic set under the saddle-node
scenario of the birth of the Smale--Williams hyperbolic attractor. In such a
mapping, a non-trivial chaotic set (with non-zero Hausdorff dimension)
arises in the general case as a result of a cascade of period-adding
bifurcations characterized by geometric scaling both in the phase space and
in the parameter space. Numerical analysis of the behavior of models
demonstrating the saddle-node scenario of birth of a hyperbolic chaotic
Smale--Williams attractor shows that these regularities are preserved in the
case of multidimensional systems. Limits of applicability of the approximate
1D model are discussed. 
\end{abstract}

\section{Introduction}

\begin{figure}[htbp]
\center{\includegraphics[width=0.9\linewidth]{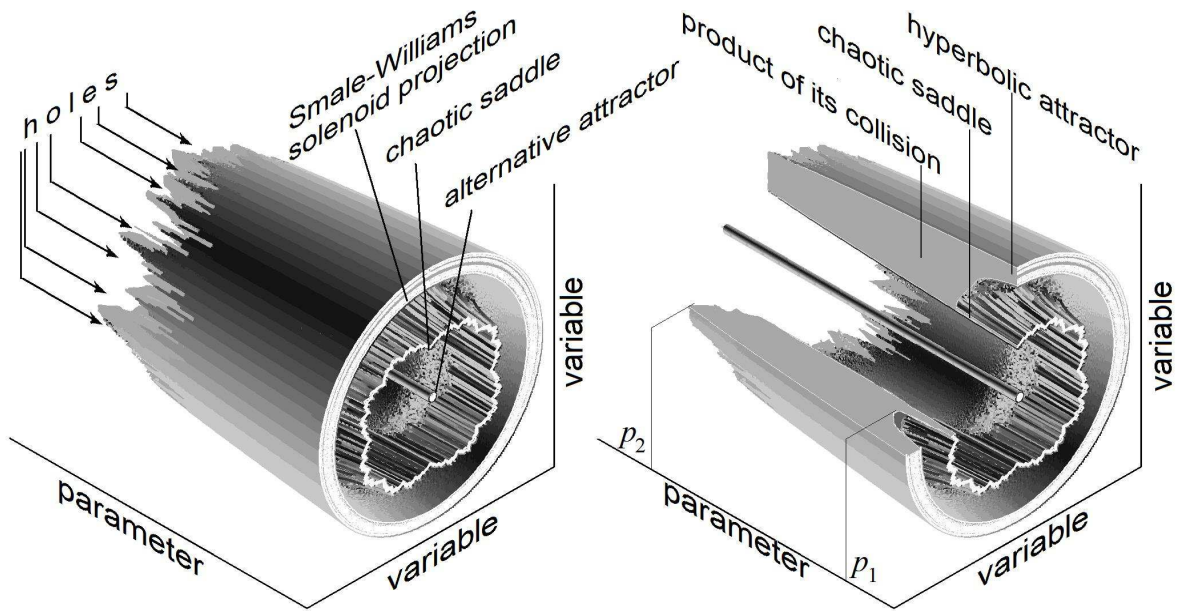}
\caption{
Scheme of the saddle-node scenario of the hyperbolic Smale--Williams 
attractor birth/destruction: 
projection of the extended space of phase variables and the control parameter is represented; 
structure, formed by the  colliding stable solenoid and saddle chaotic set, 
is shown on the left panel; 
the same structure with cut off quater part is shown on the right panel.}
\label{fig1}}
\end{figure}

\noindent The saddle-node scenario of the birth/destruction of the Smale--Williams
solenoid in a dynamical system under variation of its parameters \cite{1,2}
assumes a situation where at first the hyperbolic chaotic attractor coexists
with another attractor (possibly an infinitely distant one), and its
destruction occurs as a result of a collision/fusion with a saddle chaotic
set lying on a common boundary of their basins of attraction \cite{3,4}. In the
extended space of phase variables and the control parameter, this scenario
can be regarded as a bifurcation ``return point'' 
at which the attracting solenoid loses stability and turns into a saddle set (see 
scheme at Fig.\ref{fig1}). If we consider this
scenario when moving along the control parameter from the side where the
chaotic attractor does not yet exist, then we will observe the gradual simultaneous
birth of two chaotic sets, which eventually become a hyperbolic attractor
and a chaotic saddle. Thus, this bifurcation is prolonged and occupies a
certain interval $(p_1,p_2)$ over the control parameter. A numerical experiment shows
that the order in which the trajectories inhabiting these sets originate
depends on the parameters of the system. Which trajectories appear first, at
what point the first non-trivial chaotic set of trajectories arises from
virtually nothing, which trajectories complete the process of formation of
the chaotic attractor -- these questions need to be answered in order to
reveal the laws governing this order. The study of full-size systems is
complicated by the fact that one has to deal with highly unstable cycles. An
important role is played here by simple low-dimensional models, which allow
deep mathematical study.

In this paper, it is proposed to use the one-dimensional Bernoulli mapping
with a ''gap'' or ''hole'' to describe the Smale--Williams solenoid nascence
process \cite{10,11,12,13,14,15,16}. The mapping is well understood, which allows us to use the
results of these studies to identify the features of the saddle-node
scenario.

In the first section we show how the Bernoulli map with a hole arises in our
case. The second section briefly describes its main properties. The third
section demonstrates how the patterns characteristic of this mapping work in
the case of saddle-node scenario, and also discusses the scope of its
applicability for these purposes.

\section{Saddle-node scenario of the birth of chaotic attractor in the two-dimensional noninvertible model map}

We start with the model complex map (\ref{eq1}) introduced in [3,4]:

\begin{equation}
\label{eq1}
{z}' = \frac{Rz(z + \varepsilon )}{\sqrt {1 + \vert z(z + \varepsilon )\vert
^2} }\,\,\,.
\end{equation}

For sufficiently small values of the parameter $\varepsilon $, the dynamics
of the phase of the complex variable $z = \rho \exp (i\varphi )$ are
approximately described by the Bernoulli map, $\,\phi _{n + 1} = 2\phi _n
\,\,\,(\bmod 2\pi )$. The map (\ref{eq1}) has a stable fixed point O at the origin,
\textit{$\rho $} = 0. At values $R \approx \sqrt 2 $ and more, another attractor A appears
in the region $\rho \approx R$, which is chaotic. Taking into account the
dynamics of the phase, it can be regarded as a two-dimensional projection of
the Smale--Williams attractor.

On the common boundary of the attraction basins of both attractors there is
another chaotic set S, which is absolutely unstable. As the parameter R is
decreased, the attractor A collides with the set S and disappears, leaving
only one attractor at the origin. Thus, the mapping (\ref{eq1}) with variation of
the parameter R demonstrates a transition of the same type as the
Smale--Williams attractor under the saddle-nodal scenario.

It is known that the trajectories belonging to the attractor are encoded
using symbolic sequences of two symbols. Periodic sequences correspond to
periodic orbits. Each trajectory on the attractor corresponds to a dual
trajectory on the chaotic set S. When two sets collide, they annihilate and
disappear as a result of the saddle-node bifurcation. For different orbits,
these bifurcations occur not simultaneously, but for different values of the
parameter R. This leads to the fact that the process of collision of two
chaotic sets is distributed in a certain critical range of the parameter
values $R \in [R_1 ,\,R_2 ]$. For example, for $\varepsilon $~=~0.2, the
interval of destruction of the attractor A is determined by the values

\begin{equation}
\label{eq2}
R_{1}=1.276508545718, R_{2}= 1.482417271474.
\end{equation}

With increasing parameter $\varepsilon $ this interval widens.

\begin{figure}[htbp]
\center{\includegraphics[width=0.9\linewidth]{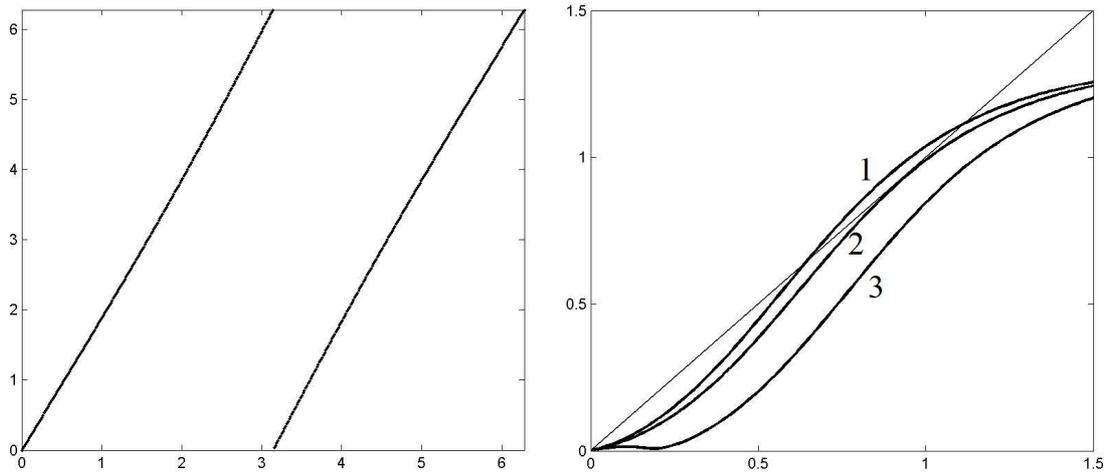}
\caption{The phase and radial components of the mapping (\ref{eq1}) at$ R$=1.35.}
\label{Fig2}}
\end{figure}

\begin{figure}[htbp]
\center{\includegraphics[width=0.8\linewidth]{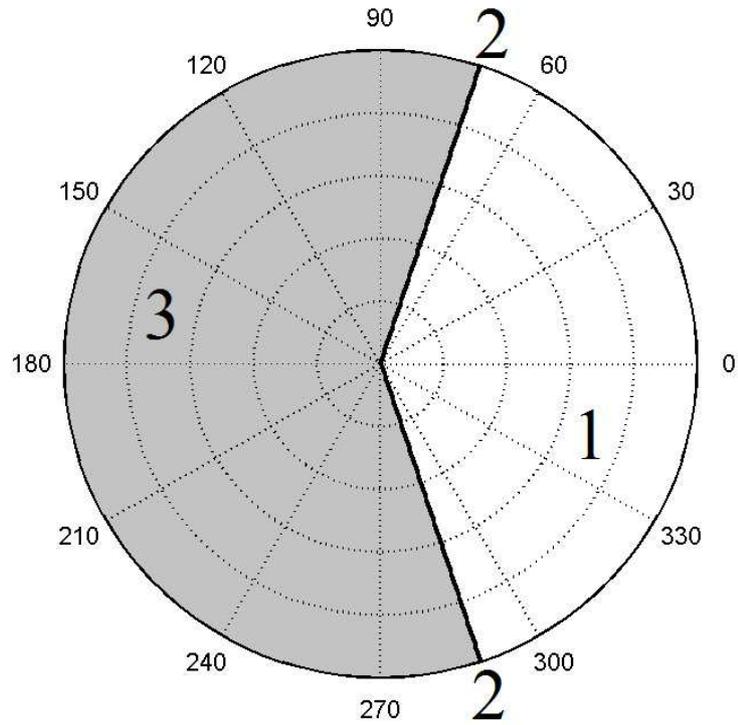}
\caption{The complex phase diagram of the mapping (\ref{eq1}) at R = 1.35. Numbers
denote areas of different behavior of the radial component of the map (\ref{eq1}),
typical graphs for each of them are presented in Fig. 2b.}
\label{Fig3}}
\end{figure}

Let us consider in more detail how the cycle of period 1 disappears. If we
take a value of R from the critical interval, then the phase dynamics
continues to be described by a Bernoulli type map and weakly depends on the
radial component over a certain range of values of $\rho $ (see Fig.~\ref{Fig2}a,
where the the graph of the complex phase component of the mapping (\ref{eq1}) is presented for $\rho
$ = 1). As for the behavior of the radial component, depending on the value
of the phase $\phi $, the map for the variable $\rho $ shows a tangent
bifurcation of the birth (or disappearance) of a pair of fixed points (see
Fig.~\ref{Fig2}b, where the graphs of the mapping are given for different values of
the phase $\phi )$. Figure~\ref{Fig3} shows a diagram in which the white color
indicates the range of phase values for which the radial dynamics
demonstrates a stable fixed point, and gray -- those for which there are no
fixed points (except that at the origin). The domain of the mapping for the
phase is split in two intervals, depending on the existence of a stable
fixed point in the mapping for the radial component. A fixed point in the
mapping (\ref{eq1}) exists only if the fixed point of the Bernoulli map for the
phase falls into the interval where there is a fixed point for the radial
component. Otherwise, the fixed point of the map (\ref{eq1}) disappears at the edge
of the gray zone as a result of the saddle-node bifurcation.

The situation can be considered in such a way that there is a ''forbidden
zone'' in phases, when a fixed point disappears into it. Obviously, the
presence of such a zone also affects the stability of longer-period cycles
-- if sufficiently many points of the orbits of the Bernoulli mapping fall
into a forbidden zone, then such a cycle in the map (\ref{eq1}) also disappears. If
we extremely simplify the situation and assume that any orbit disappears, at
least one point of which falls into the forbidden band, then we arrive at
the simplest one-dimensional model that describes disappearance of the
trajectories under the saddle-node scenario -- the Bernoulli map with a
''forbidden zone'' or ''hole''.

\section{A model in the form of a Bernoulli mapping with a hole}

The mapping $T$ of the Bernoulli shift with the forbidden interval of values of
the variable, when hit in which the trajectory is considered impossible (in
real models this corresponds to running away to another attractor, possibly
at infinity)

\begin{equation}
\label{eq3}
x_{n + 1} = \left\{ {{\begin{array}{*{20}c}
 {2x_n \,\,(\bmod \,\,1),\,\,\,x_n \notin {\rm H}^0,} \hfill \\
 {not\;defined,\,\,otherwise,} \hfill \\
\end{array} }} \right. {\rm where\,}H^{0}=[a+b/2,a-b/2].
\end{equation}

The parameters $a$ and $b$ here are responsible for the position of the center and
the width of the hole, respectively.\footnote{ Models in the form of
mappings with a hole and leakage were investigated earlier in \cite{10,11,12,13,13a,13b,13c,13d}. It was shown that
the average ''lifetime'' of trajectories in such systems depends in a complex
manner on the position of the band gap.}

\begin{figure}[htbp]
\center{\includegraphics[width=0.6\linewidth]{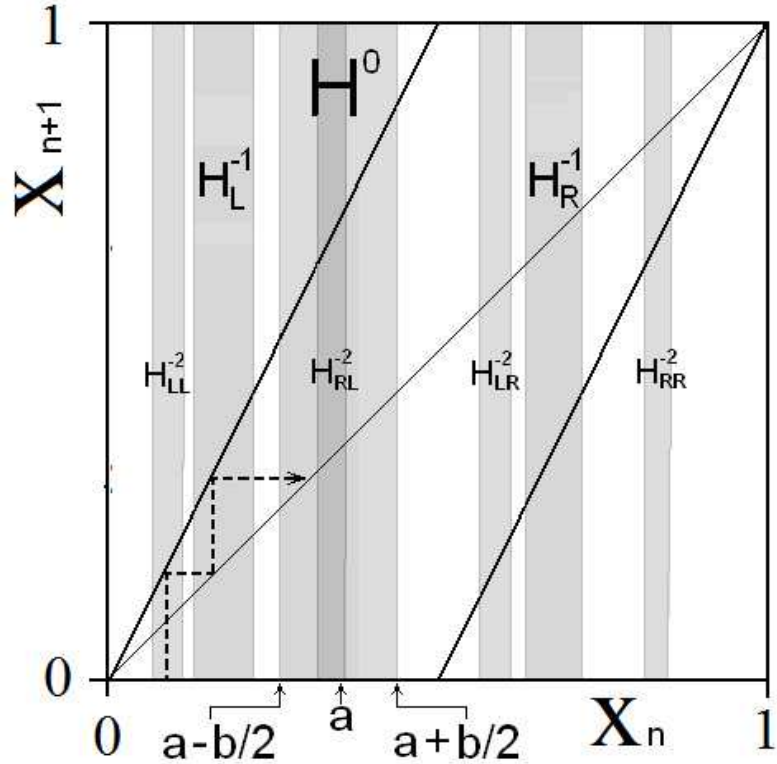}
\caption{An iterative diagram of the Bernoulli map with a ''hole'' H$^{0}$ (gray
strip) with center at a and width b. Several inverse images of the forbidden
band are also depicted.}
\label{Fig4}}
\end{figure}

From the point of view of analogy with the system (\ref{eq1}), the parameter $b$ can be
associated with the value $\vert $\textit{R$_{2}\vert $}--$\vert $\textit{R$\vert $}, that is, the degree of deepening
into the critical interval. The parameter $a$ can be in turn associated with
the argument of R, assuming this parameter is a complex value.

The set C of points whose orbits do not enter the hole, ${\rm C(a, b) :} =
 \{{\rm x } \in {\rm [0, 1] : T}^{\rm n}{\rm x} \notin {\rm (a, b), n }
\ge {\rm 0}\}$, can be defined as the limit

\begin{equation}
\label{eq4}
{\rm C} = \mathop {\lim }\limits_{k \to \infty } {\rm C}_k
,\,\,\,\,\,\,\,{\rm C}_k = [0,1]\backslash \bigcup\nolimits_{j = 0}^k
{\rm H}^{ - j},
\end{equation}

\noindent
where H$^{ - j}$ is the union of all the jth inverse images of the original
hole H$^{0}$ with all possible encodings (see the diagram in Fig.~\ref{Fig4}, where
two preimages are given).

\begin{figure}[htbp]
\center{\includegraphics[width=0.9\linewidth]{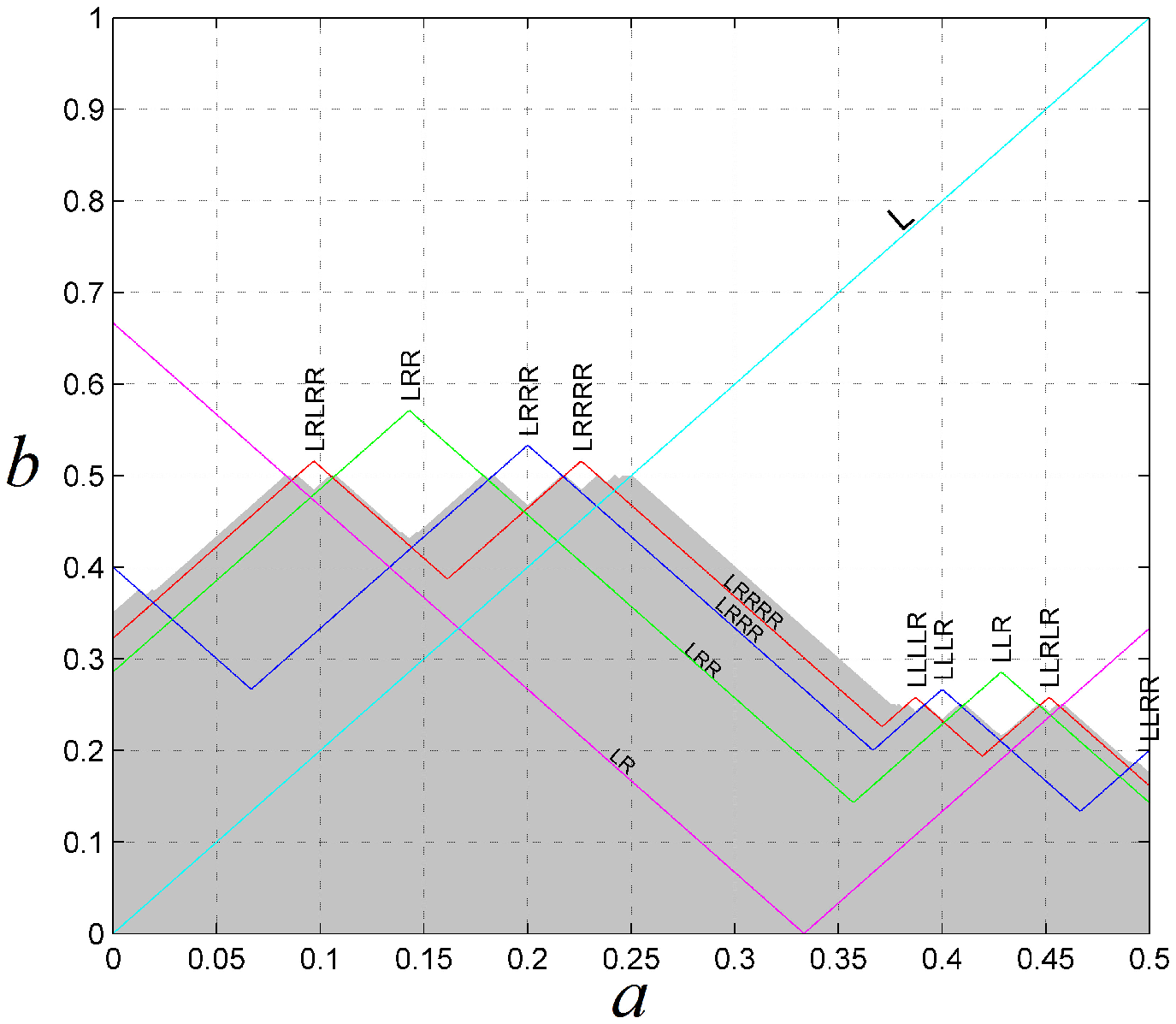}
\caption{The lines are the borders of the domains where at least one cycle
exists of a given period. Encodings of the cycles which are born at the
borders are shown. The domain \textbf{D }is highlighted in gray.}
\label{Fig5}}
\end{figure}

Fig.~\ref{Fig5} shows the arrangement of the ''position -- width'' parameter plane. The
lines correspond to the boundaries of the existence of cycles of different
periods. Below each of these boundaries, there is at least one cycle of the
corresponding period. Symbols above the lines denote the symbolic code of
the cycle that arises first of all the cycles of a given period on this
section of the boundary

We define \textbf{D} as a set of points on the plane of the parameters ($a$,
$b)$ for which the Hausdorff dimension of the set C is not zero. In Fig.~\ref{Fig5},
the numerically calculated set \textbf{D} is highlighted in gray. The
picture occurs to be symmetrical and only half of it is presented.

In \cite{14} it was proved that the boundary of this set consists of points of
three types: \newline
1. the point at which from all non-periodic trajectories the first appears
an orbit with a quasi-periodic symbolic code corresponding to irrational
rotations of the circle maps \cite{17}; \newline
2. the point at which from all non-periodic trajectories the first appears
an orbit with a symbolic code corresponding to tupling (doubling, tripling
...) of the period; \newline
3. the point at which out of all the non-periodic trajectories the first
appears an orbit with symbolic code B{\{}À{\}}$^{\infty }$, where A and B
are some combinations of R and L's, such an orbit is often called
pre-periodic to a certain cycle.

The transition through this boundary corresponds to appearance of a
nontrivial chaotic set, so that this can be considered as a transition to
chaos. Theorem proved in \cite{14}, states that this transition, depending on the
point at which we cross the border can be, as stated above, the one of three
possible types -- via ''quasiperiodicity'', via period tupling, and via the
occurrence of the preperiodic trajectory. The codimension of the first two
types of points is equal to 2, they are isolated points on the parameter
plane. Codimension of the set of the third type of critical points of
transition to chaos is 1, that is, there are segments of lines consisting of
such points on the parameter plane. (To be precise, we may add that
codimension of the critical behavior of this type may be also 2. This
corresponds to points at the ends of the segments). To these lines from
below the lines are accumulating of appearance of cycles, symbolic codes of
which correspond to the sequence of period adding. And from all cycles of
the corresponding period $N$ and less, the cycles with such symbolic codes will
appear before the others. Thus, in the general case, when parameters are
varied following along a certain path, a non-trivial chaotic set will arise
through the sequence of occurrence of cycles, the symbolic codes of which
correspond to the sequence of period adding.

\begin{figure}[htbp]
\center{\includegraphics[width=0.9\textwidth]{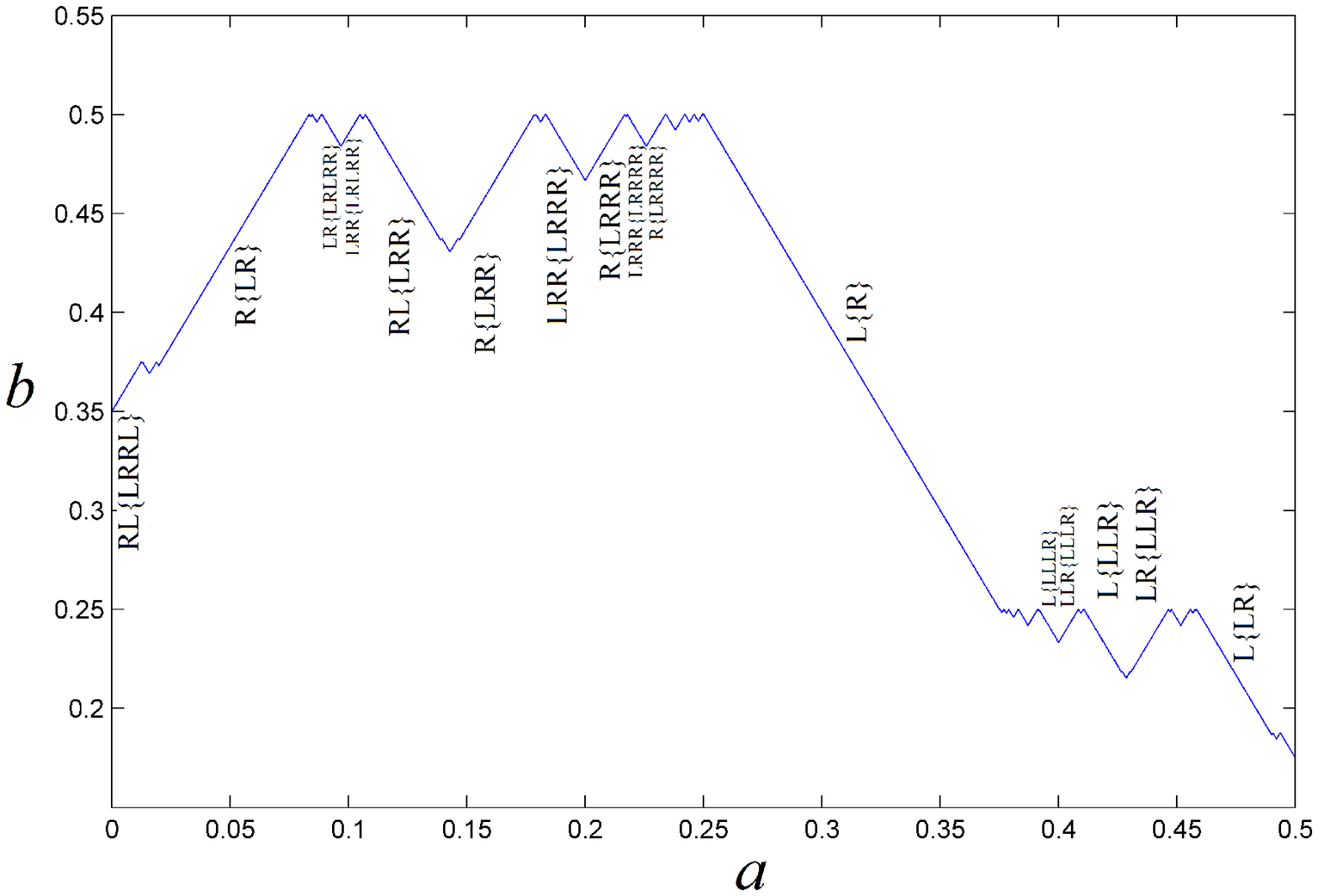}
\caption{The boundary of the domain \textbf{D}. Symbolic codes of non-periodic
trajectories arising on it are indicated for some segments.}
\label{Fig6}}
\end{figure}

Figure~\ref{Fig6} shows the set \textbf{D} whose boundary was constructed this time as the
upper bound of the set of lines of occurrence of cycles with symbolic
sequences of the form B{\{}À{\}}$^{n}$, where n denotes the n-fold
repetition of the symbol A, and the symbols A and B themselves denote all
possible sequences characters R and L of finite length m$ \le $6. For some
sections of the boundary, symbolic codes of the non-periodic trajectories
arising on it are indicated.

\begin{figure}[htbp]
\center{\includegraphics[width=0.9\linewidth]{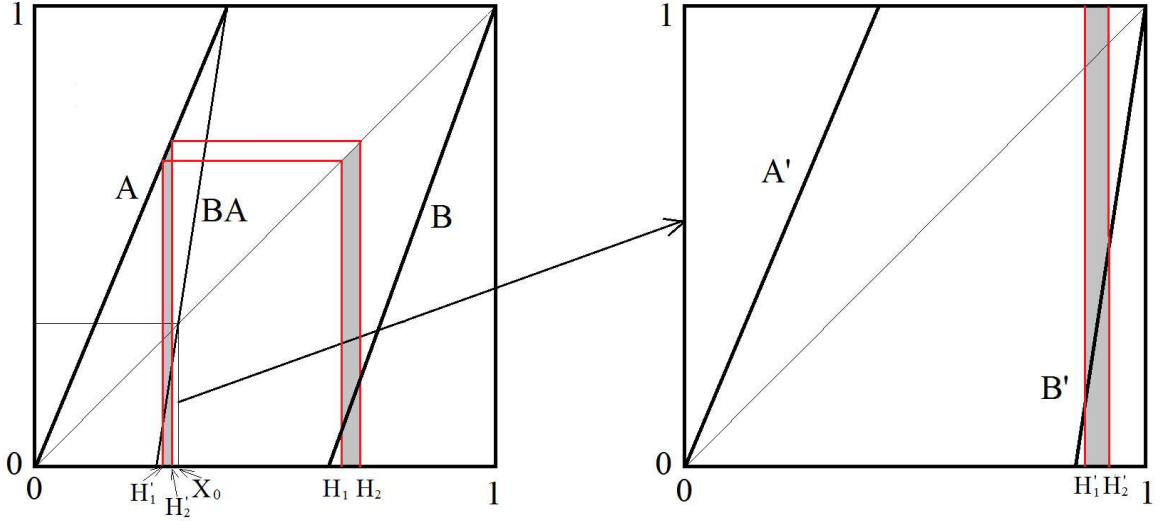}
\caption{Renormalization scheme for the period adding case.}
\label{Fig7}}
\end{figure}

In order to clarify the quantitative laws of accumulation of the lines of
occurrence of cycles to the boundary of the set \textbf{D}, we consider a
simple renormalization scheme presented in Fig.~\ref{Fig7}. The ''inflation'' rule \cite{17}
in the case of a sequence of period adding can be written in the form

\begin{equation}
\label{eq5}
\left\{ {\begin{array}{l}
 {\rm {\bf A}} \\
 {\rm {\bf B}} \\
 \end{array}} \right. \Rightarrow \left\{ {\begin{array}{l}
 {\rm {\bf A}} \\
 {\rm {\bf BA}} \\
 \end{array}} \right.,
\end{equation}

The symbols A and B correspond to segments of graphs of linear functions

\begin{equation}
\label{eq6}
\left\{ {\begin{array}{l}
 {\rm {\bf A}}(x) \\
 {\rm {\bf B}}(x) \\
 \end{array}} \right. = \left\{ {\begin{array}{l}
 \frac{x}{K_L } \\
 \frac{x - 1}{K_R } + 1 \\
 \end{array}} \right.,
\end{equation}

The origin corresponds to a cycle with a symbolic sequence {\{}A{\}}. The
intersection point of the graph of the composite function \textbf{BA} with
the diagonal, ${\rm x}_{\rm 0} = \frac{K_L (1 - K_R )}{(1 - K_R K_L )}$,
corresponds to the cycle point with the corresponding symbolic code BA.
Obviously, if we iterate the map (\ref{eq3})$ n$ times, we get as a result a point
belonging to the orbit with the desired symbolic code B{\{}À{\}}$^{n}$.

After rescaling the phase variable in \textbf{$\alpha $}=1/x$_{0}$, we
obtain for new functions ${\rm {\bf A}}'$ and ${\rm {\bf B}}'$

\begin{equation}
\label{eq7}
\left\{ {\begin{array}{l}
 {\rm {\bf A}}'(x) \\
 {\rm {\bf B}}'(x) \\
 \end{array}} \right. = \left\{ {\begin{array}{l}
 \frac{x}{K'_L } \\
 \frac{x - 1}{K'_R } + 1 \\
 \end{array}} \right.,
\end{equation}

\noindent
where

\begin{equation}
\label{eq8}
\begin{array}{l}
  K'_L {\rm = }K_L   \\
 K'_R = K_L K_R \\
 \end{array},
\end{equation}

For expanding maps of Bernoulli type $K_L ,\,K_R < 1$, therefore, in the
limit of a large number of iterations $K_R \to 0$, and the scale factor is
$\alpha \to \frac{1}{K_L }$. It follows that the distance from the nearest
element of the cycle B{\{}À{\}}$^{n}$ to the cycle A is subject to geometric
scaling with the scale factor \textbf{$\alpha $}=1/$K_{L}$. Hence, actually
it is determined by the slope of that branch of the map to which the cycle
of the adding period belongs.

It should be noted that this is simply a property of Bernoulli-type
mappings, since we have not yet introduced the presence of a forbidden zone.
Let us now consider the dynamics of a hole.

It can be seen from Fig.~\ref{Fig7} that the new positions of the edges of the hole
are determined from the relation ${\rm A(H'}_{{\rm 1,2}} {\rm )} = {\rm
H}_{{\rm 1,2}} $. After the renormalization, we obtain

\begin{equation}
\label{eq9}
{\rm H'}_{{\rm 1,2}} {\rm = H}_{{\rm 1,2}} K_L {\rm / x}_{\rm 0} = {\rm
H}_{{\rm 1,2}} \frac{(1 - K_R )}{(1 - K_R K_L )},
\end{equation}

The parameters $a$ and $b$ are expressed in terms of H$_{1,2}$ as

\begin{equation}
\label{eq10}
a = \frac{{\rm H}_{\rm 1} + {\rm H}_{\rm 2} }{{\rm 2}},\quad b = {\rm
H}_{\rm 2} - {\rm H}_{\rm 1} ,
\end{equation}

\noindent
then for them the following relations are also valid

\begin{equation}
\label{eq11}
 {a}' = a\frac{(1 - K_R )}{(1 - K_R K_L )}\quad b{\rm '} = b\frac{(1 -
K_R )}{(1 - K_R K_L )},
\end{equation}

The factor $ \frac{(1 - K_R )}{(1 - K_R K_L )}$ in the limit tends to
1, so the scaling factor for parameters $a$ and $b$ is the same as for the other
objects in the phase space, $\delta =\alpha $=1/$K_{L}$.

As an example, consider emergence of a chaotic set as we reduce the hole
width \textbf{\textit{b}}, while the position of its center (the value of
parameter b) does not change. We take two values of the parameter $a$:
$a_{1}$=0.3 and $a_{2}$=0.075, the first belongs to the longest rectilinear
portion of the boundary \textbf{$\Gamma $} corresponding to the addition of
period 1, and the second to the section where the addition of period 2 is
observed. In both cases, all values of the parameter $a$ were calculated, which
correspond to the birth of cycles with periods corresponding to all possible
symbolic codes of length $ \le $20.

In the first case, at $a$ = 0.3, it was found that of all cycles with periods
equal to or less than $N$, the first cycle to born is that with a symbolic code
of the form LR$^{N - 1}$. Table 1 lists the values of the parameter $a$, for
which the cycles with such codes appear, as well as the elements of these
cycles closest to the cycle of period 1 with the code {\{}R{\}}. In 
Figure~\ref{Fig5}, one can see how the lines of occurrence of cycles with codes of the form
LR$^{N - 1}$ accumulate to the boundary \textbf{$\Gamma $} of the domain
\textbf{D}. The scale factors \textbf{$\delta $} and \textbf{$\alpha $} are
set by the slope of the R branch of the map and are equal to 2.

\begin{table}[h!]
\doublerulesep 0.1pt
\tabcolsep 7.8mm
\centering
\caption{\rm The sequence of bifurcation points of cycles with the codes LR$^{N -
1}$ of the map (\ref{eq3}) at \textbf{a}=0.3.}
\vspace*{2mm}
\renewcommand{\arraystretch}{1.3}
\setlength{\tabcolsep}{17pt}
\footnotesize{\begin{tabular*}{16.5cm}{cccccc}
\hline\hline\hline
N&
\textbf{Code}&
$b_{N}$ &
$x_{N}$&
$\frac{(b_{N - 2}-b_{N - 1})}{(b_{N - 1}-b_{N})}$&
$\frac{(x_{0}-x_{N - 1})}{(x_{0}-x_{N})}$ \\
\hline
1&
LR&
0.066666667&
0.66666667&
&
 \\
2&
LRR&
0.25714286&
0.85714286&
&
2.3333333 \\
3&
LRRR&
0.33333333&
0.93333333&
2.5000000&
2.1428571 \\
4&
LRRRR&
0.36774194&
0.96774194&
2.2142857&
2.0666667 \\
5&
LRRRRR&
0.38412698&
0.98412698&
2.1000001&
2.0322581 \\
6&
LRRRRRR&
0.39212598&
0.99212598&
2.0483866&
2.0158730 \\
7&
LRRRRRRRR&
0.39607843&
0.99607843&
2.0238101&
2.0078740 \\
8&
LRRRRRRRRR&
0.39804305&
0.99804305&
2.0118117&
2.0039216 \\
9&
LRRRRRRRRRR&
0.39902248&
0.99902248&
2.0058793&
2.0019569 \\
10&
LRRRRRRRRRRR&
0.39951148&
0.99951148&
2.0029430&
2.0009775 \\
\ldots &
&
\ldots &
\ldots &
\ldots &
\ldots  \\
$\infty $&
LR$^{\infty }$&
0.4&
1.0&
2.0&
2.0 \\
\hline\hline\hline\end{tabular*}
}
\renewcommand{\arraystretch}{1}
\label{tab1}
\end{table}

In the second case, at $a$ = 0.075, it was found that out of all cycles with
periods less than 2*N+2, the first appears a cycle, with a symbolic code of
the form R{\{}RL{\}}$^{N}$. Table 2 shows the values of the parameter $b$, for
which the cycles with such codes appear, as well as the elements of these
cycles closest to the element of the cycle of period 2 with the code
{\{}RL{\}}. The scaleing factors $\delta $ and $\alpha $ in this case are
given by the slope of the RL branch of the doubly iterated map and are equal
to 4.

\begin{table}[h!]
\doublerulesep 0.1pt
\tabcolsep 7.8mm
\centering
\caption{\rm The sequence of bifurcation points of cycles with codes R{\{}RL{\}}$^{N}$ of the map (\ref{eq3}) at \textbf{a}=0.075.}
\vspace*{2mm}
\renewcommand{\arraystretch}{1.3}
\setlength{\tabcolsep}{17pt}
\footnotesize{\begin{tabular*}{16.5cm}{cccccc}
\hline\hline\hline
N&
Code&
$b_{N}$&
$x_{N}$&
$\frac{(b_{N - 2}-b_{N - 1})}{(b_{N - 1}-b_{N})}$&
$\frac{(x_{0}-x_{N - 1})}{(x_{0}-x_{N})}$ \\
\hline
1&
RRL&
0.43571428&
0.85714286&
&
 \\
2&
RRLRL&
0.47258065&
0.70967742&
&
4.4285714 \\
3&
RRLRLRL&
0.48070866&
0.67716535&
4.5357139&
4.0967742 \\
4&
RRLRLRLRL&
0.48268102&
0.66927593&
4.1209703&
4.0236220 \\
5&
RRLRLRLRLRL&
0.48317049&
0.66731803&
4.0295260&
4.0058708 \\
6&
R{\{}RL{\}}$^{6}$&
0.48329264&
0.66682945&
4.0073383&
4.0014656 \\
7&
R{\{}RL{\}}$^{7}$&
0.48332316&
0.66670736&
4.0016906&
4.0003663 \\
8&
R{\{}RL{\}}$^{8}$&
0.48333079&
0.66667684&
4.0007515&
4.0000916 \\
9&
R{\{}RL{\}}$^{9}$&
0.48333270&
0.66666921&
4.0000000&
4.0000229 \\
10&
R{\{}RL{\}}$^{10}$&
0.48333317&
0.66666730&
4.0080321&
4.0000057 \\
\ldots &
&
&
&
&
 \\
$\infty $&
R{\{}RL{\}}$^{\infty }$&
0.4833333333&
0.66666666&
4.0&
4.0 \\
\hline\hline\hline\end{tabular*}
}
\renewcommand{\arraystretch}{1}
\label{tab2}
\end{table}

\bigskip

\section{Two-dimensional noninvertible mapping}

In the case of two-dimensional model map (\ref{eq1}), with the values of the
parameters from Section 1, the numerical analysis shows that here the case
of adding a cycle of period 1 is realized with the appearance on the
boundary \textbf{$\Gamma$} of a non-periodic trajectory with the code RL$^{\infty
}$. In this case, from all cycles with periods equal to or less than $N$, the
first cycle with a symbolic code of the form RL$^{N - 1}$ occurs. Table 3
lists the corresponding bifurcation values of the parameter R$_{N}$, as well
as the $x_{N}$ and $y_{N}$ coordinates of the elements of these cycles at
tangential bifurcation points closest in phase (which is defined as $\varphi _N = \arg \,z_N /
2\pi $), to the cycle of period 1.

\begin{table}[h!]
\doublerulesep 0.1pt
\tabcolsep 7.8mm
\centering
\caption{\rm The sequence of bifurcation points of cycles with codes
RL$^{N}$ of the map (\ref{eq1}) at $\varepsilon $=0.2.}
\vspace*{2mm}
\renewcommand{\arraystretch}{1.3}
\setlength{\tabcolsep}{11pt}
\footnotesize{\begin{tabular*}{16.5cm}{ccccccc}
\hline\hline\hline
N&
Code&
R$_{n}$&
Re z$_{n}$&
Im z$_{n}$&
$\alpha_N = \frac{(\varphi _{N - 2} - \varphi _{N - 1} )}{(\varphi _{N - 1} - \varphi _N )}$ &
$\delta_N = \frac{(R_{N - 2} - R_{N - 1} )} {(R_{N - 1} - R_N )}$  \\
\hline
&
L&
1.2765086&
0.854011633&
0.0&
&
 \\
1&
RL&
1.4824173&
-0.58795349&
8.9548117e-01&
&
 \\
2&
RLL&
1.4429370&
0.50695195&
8.3766455e-01&
&
 \\
3&
RLLL&
1.4061755&
0.78301132&
4.5912542e-01&
-4.0633411&
2.3402559 \\
4&
RL$^{4}$&
1.3819309&
0.82699493&
2.4067062e-01&
2.0079764&
2.1024585 \\
5&
RL$^{5}$&
1.3657408&
0.82033001&
1.2707420e-01&
1.9081841&
1.9583017 \\
6&
RL$^{6}$&
1.3544064&
0.80515059&
6.7879207e-02&
1.8614063&
1.8620736 \\
7&
RL$^{7}$&
1.3461006&
0.79074973&
3.6591647e-02&
1.8374772&
1.7742608 \\
8&
RL$^{ 8}$&
1.3397788&
0.77870708&
1.9852908e-02&
1.8246373&
1.6702997 \\
9&
RL$^{ 9}$&
1.3348195&
0.76887703&
1.0819726e-02&
1.8175212&
1.5445921 \\
10&
RL$^{ 10}$&
1.3308354&
0.76083082&
5.9155441e-03&
1.8134651&
1.4167726 \\
11&
RL$^{ 11}$&
1.3275739&
0.75417882&
3.2418203e-03&
1.8110817&
1.3143631 \\
12&
RL$^{ 12}$&
1.3248635&
0.74861897&
1.7797214e-03&
1.8096277&
1.2470655 \\
13&
RL$^{ 13}$&
1.3225838&
0.74392519&
9.7839684e-04&
1.8086976&
1.2073214 \\
14&
RL$^{ 14}$&
1.3206473&
0.73992816&
5.3846671e-04&
1.8080671&
1.1841555 \\
15&
RL$^{ 15}$&
1.3189892&
0.73649947&
2.9661763e-04&
1.8076110&
1.1698914 \\
16&
RL$^{ 16}$&
1.3175601&
0.73354019&
1.6351765e-04&
1.8072585&
1.1603315 \\
17&
RL$^{ 17}$&
1.3163215&
0.73097293&
9.0201188e-05&
1.8069693&
1.1533936 \\
18&
RL$^{ 18}$&
1.3152432&
0.72873623&
4.9785269e-05&
1.8067201&
1.1480667 \\
19&
RL$^{ 19}$&
1.3143010&
0.72678061&
2.7491549e-05&
1.8064974&
1.1438448 \\
20&
RL$^{ 20}$&
1.3134753&
0.72506577&
1.5187344e-05&
1.8062930&
1.1404535 \\
\hline\hline\hline\end{tabular*}
}
\renewcommand{\arraystretch}{1}
\label{tab3}
\end{table}

These components of the cycles accumulate in two-dimensional phase space to
certain points, which in this case do not longer belong to the cycle of
period 1. However, if we consider an approximate one-dimensional map for the
phase, then it turns out that the point of accumulation of the phase
components is the point corresponding to the phase of the cycle of period
one. Table 3 gives numerical estimates for the corresponding scale factor
$\alpha  \approx $1.806. A good approximation for the slope of the branch
L at the accumulation point can be obtained from the values of the
multipliers of the cycle of period one at the bifurcation point. The
multiplier, which is equal to one, corresponds to the tangent bifurcation in
the radial dynamics, while the value of the leading multiplier characterizes
the phase component of the dynamics and gives a good estimate for the slope
coefficient of the corresponding branch of the approximate one-dimensional
mapping for the phase, and hence for the phase scale factor, $\alpha
$=1/K$_{L} \approx \mu _1 $=1.81.

The sequence of bifurcation points of the corresponding cycles with respect
to the parameter \textbf{R} demonstrates geometric convergence. The
numerical estimate for the scaling factor in the parameter space is $\delta
 \approx $1.14, which differs from that predicted by the model $\delta
$=1/K$_{L} \approx \mu _1 $=1.81. The reason for this is obviously that
the assumption made in the derivation of the model mapping that the cycle
disappears if at least one of the elements of its orbit falls into a hole,
is too strong. In fact, of course, the cycle disappears only after a
sufficiently large number of its elements are in the forbidden zone, which
leads to a protraction of the cascade.

\bigskip

\section{Saddle-node scenario in the 4D invertible model map}

As a second example of a system demonstrating the scenario of interest, we
consider a reversible four-dimensional system with discrete time, which was
introduced in \cite{4}. Such a model is constructed on the basis of a map with a
complex variable (\ref{eq1}) in strict analogy to the construction of the Henon map:

\begin{equation}
\label{eq12}
z_{n + 1} = \frac{Rz_n (z_n + \varepsilon )}{\sqrt {1 + \vert z_n (z_n +
\varepsilon )\vert ^2} } - bz_{n - 1} ,
\end{equation}

The attractor is now located in the four-dimensional space and is a
full-fledged Smale--Williams solenoid. We consider the dynamics of this
mapping for the values of the parameters $\varepsilon $~= 0, b = 0.3. The
saddle-node scenario of disappearance of the chaotic attractor is still
observed with decreasing the parameter $R$.

Numerical analysis shows that here the case of cycle of period 2 adding is
realized with the appearance on the boundary of a nonperiodic trajectory
with the code R{\{}RL{\}}$^{\infty }$. Of all cycles with periods less than
2$N$+2, the first a cycle appears, with symbolic code of the form
R{\{}RL{\}}$^{N}$. Table 4 shows the bifurcation values of the parameter
R$_{N}$ and the phase \textit{$\phi $}$_{N}$ of the closest (in phase) cycle components to
one of the elements of the cycle of period 2.

An estimate of the average rate of convergence in phase can be obtained from
the largest multiplier of the cycle of period 2 at the point of its
disappearance \textbf{$\alpha $}=1/K$_{L} \approx \mu _1 $=2.5. The
numerical value of the scale factor over the parameter K is again
significantly lower than the value assumed by the model.

\begin{table}[h!]
\doublerulesep 0.1pt
\tabcolsep 7.8mm
\centering
\caption{\rm The sequence of bifurcation points of cycles with codes
R{\{}RL{\}}$^{N}$ of the map (\ref{eq12}) at $\varepsilon $ = 0, b = 0.3.}
\vspace*{2mm}
\renewcommand{\arraystretch}{1.3}
\setlength{\tabcolsep}{14pt}
\footnotesize{\begin{tabular*}{16.5cm}{cccccc}
\hline\hline\hline
N&
code&
\textit{$\phi $}$_{N}$ &
R$_{N}$&
$\alpha _N = \frac{(\varphi _{N - 2} - \varphi _{N - 1} )} {(\varphi _{N - 1} - \varphi _N )}$&
$\delta _N = \frac{(R_{N - 2} - R_{N - 1} )} {(R_{N - 1} - R_N )}$ \\
\hline
2&
RL&
0.33333333&
0.9899494937&
&
 \\
3&
RRL&
0.46357761&
1.5162512&
&
 \\
5&
RRLRL&
0.37897213&
1.3678263&
&
 \\
7&
RRLRLRL&
0.34971544&
1.3004457&
2.8918343&
2.2027843 \\
9&
RRLRLRLRL&
0.33845392&
1.2657087&
2.5979340&
1.9397398 \\
11&
RRLRLRLRLRL&
0.33390738&
1.2461462&
2.4769408&
1.7756890 \\
13&
R{\{}RL{\}}$^{6}$&
0.33197756&
1.2345261&
2.3559391&
1.6835022 \\
\hline\hline\hline\end{tabular*}
}
\renewcommand{\arraystretch}{1}
\label{tab4}
\end{table}

\section{The model of two coupled Van der Pol oscillators}

It was shown in \cite{3} that the considered scenario is realized in the system
of two coupled van der Pol oscillators \cite{5}

\begin{equation}
\label{eq13}
\begin{array}{l}
 \dot {x} = \omega _0 u,\,\,\,\,\,\,\,\,\,\, \\
 \dot {u} = (h + a\cos {2\pi t} \mathord{\left/ {\vphantom {{2\pi t} T}}
\right. \kern-\nulldelimiterspace} T - x^2)u - \omega _0 x + (\varepsilon y
/ \omega _0 )\cos \omega _0 t, \\
 \dot {y} = 2\omega _0 v,\,\,\,\,\,\,\, \\
 \dot {v} = (h - a\cos {2\pi t} \mathord{\left/ {\vphantom {{2\pi t} T}}
\right. \kern-\nulldelimiterspace} T - y^2)v - 2\omega _0 y + (\varepsilon
\mathord{\left/ {\vphantom {\varepsilon {2\omega _0 }}} \right.
\kern-\nulldelimiterspace} {2\omega _0 })x^2, \\
 \end{array}
\end{equation}

The hyperbolicity of the attractor at the values of the parameters $h$=0,
$\omega _{0}$=2$\pi $, $T$=6, $A$=5, $\varepsilon $=0.5 was tested numerically
using the cone criterion \cite{6,7}, and it was verified using the
computer-assisted proof framework \cite{8}. On the parameter plane ($h$, $A)$, for
negative $h$, the region of existence of the hyperbolic chaotic attractor is
bounded by the line on which this attractor arises as a result of the
saddle-node scenario.

In \cite{3}, the creation of a chaotic attractor with variation of the parameter
$h$ for a constant $A$~= 6.5 was considered. Along this path, the cycle of period
3 with code LLR appears first. The code of the cycle was determined from the
mapping for the phase, which was calculated as $\varphi _n = \arg [x_0 (t_n
) + ix_1 (t_n )] / 2\pi $, where $\{x_0 ,x_1 \} = \{x,\,\,u / 0.9 - x / 2\}$
is the variable change in the original equation (\ref{eq13}), made in order that in
the new coordinates the attractor was close to the circle, and
$t_{n}$\textit{=nT}, which corresponds to the mapping in the stroboscopic section. As the
parameter $h$ is increased, cycles with a period less than or equal 10 start
with cycles LRLLR and LRLLRLLR, indicating that there is a sequence of
addition of period 3. However, finding out the bifurcation points of cycles
of a larger period is more difficult because of the fact that they quickly
become strongly unstable.

If one moves up the parameter $A$, then at $A$ = 7.1, the cycle of period 2 is the
first to appear with increase of the parameter $h$, which allows us to hope
that there will be a scenario of adding this period. Indeed, numerical
analysis shows that of all cycles with periods less than 2$N$+2, the first
cycle appears with a symbolic code of the form L{\{}LR{\}}$^{N}$. Table 5
shows values of the parameter $h_{N}$, for which cycles with such codes arise
as a result of the tangent bifurcation, as well as phase values $\varphi _n
$ for the elements of these cycles are closest to one of the elements of the
cycle of period 2. The estimate for the scaling factor determining the
accumulation rate for the phases $\varphi _N $ can be obtained from the
value of the largest multiplier of the cycle of period 2 at the moment of
disappearance as a result of tangent bifurcation, 1/K$_{L} \approx \mu
_n $=3.99.

\begin{table}[h!]
\doublerulesep 0.1pt
\tabcolsep 7.8mm
\centering
\caption{\rm The sequence of bifurcation points of cycles with codes
L{\{}LR{\}}$^{N}$ of the system (\ref{eq13}) at $A$ = 7.1.}
\vspace*{2mm}
\renewcommand{\arraystretch}{1.3}
\setlength{\tabcolsep}{14pt}
\footnotesize{\begin{tabular*}{16.5cm}{cccccc}
\hline\hline\hline
N&
code&
$h_{N}$ &
$\varphi _N $&
$\alpha _N = \frac{(\varphi _{N - 2} - \varphi _{N - 1} )}{(\varphi _{N - 1} - \varphi _N )}$&
$\delta _N = \frac{(h_{N - 2} - h_{N - 1} )}{(h_{N - 1} - h_N )}$ \\
\hline
&
LR&
-1.5784956&
0.30920170&
&
 \\
1&
LLR&
-1.5782312&
0.26073326&
&
 \\
2&
LLRLR&
-1.5784217&
0.29818979&
&
 \\
3&
LLRLRLR&
-1.5784563&
0.30643015&
4.5454967&
5.4964218\\
4&
LLRLRLRLR&
-1.5784680&
0.30840628&
4.1699353&
2.9684770 \\
\hline\hline\hline\end{tabular*}
}
\renewcommand{\arraystretch}{1}
\label{tab5}
\end{table}

On the other hand, if we consider the boundary of the region of existence of
a chaotic attractor for small negative constant $h$ and vary the parameter $A$,
then at least for cycles of a not very long period, the addition of period 1
may be observed, see Table 6. However, we should pay attention to that
circumstance that the cycle of period one arises here only after the cycle of
period 9 have already appeared, that is, it does not exist at a
possible point of accumulation of bifurcation points. Moreover, the cycle of
period one, in contrast to all other cycles, becomes stable at the
bifurcation point.

\begin{table}[h!]
\doublerulesep 0.1pt
\tabcolsep 7.8mm
\centering
\caption{\rm The sequence of bifurcation points of cycles with codes RL$^{N}$
of the system (\ref{eq13}) at h = -0.05.}
\vspace*{2mm}
\renewcommand{\arraystretch}{1.3}
\setlength{\tabcolsep}{14pt}
\footnotesize{\begin{tabular*}{16.5cm}{cccccc}
\hline\hline\hline
N&
code&
$A_{N}$ &
$\varphi _N $&
$\alpha _N = \frac{(\varphi _{N - 2} - \varphi _{N - 1} )}{(\varphi _{N - 1} - \varphi _N )}$&
$\delta _N = \frac{(A_{N - 2} - A_{N - 1} )}{(A_{N - 1} - A_N )}$ \\
\hline
&
R&
3.9307609&
0.0&
&
 \\
1&
RL&
4.6171255&
0.73543964&
&
 \\
2&
RLL&
4.3823356&
0.49955719&
&
 \\
3&
RLLL&
4.2265313&
0.30150912&
1.1910364&
1.5069537 \\
4&
RL$^{4}$ &
4.1254594&
0.22563264&
2.6101379&
1.5415202 \\
5&
RL$^{5}$ &
4.0516771&
0.18146253&
1.7178240&
1.3698668 \\
6&
RL$^{6}$ &
3.9909290&
0.14728899&
1.2925235&
1.2145612 \\
7&
RL$^{7}$ &
3.9379366&
0.11646029&
1.1084981&
1.1463548 \\
8&
RL$^{ 8}$ &
3.8929254&
0.086644303&
1.0339650&
1.1773160 \\
9&
RL$^{ 9}$ &
3.8616612&
0.057894430&
1.0370825&
1.4396996 \\
\hline\hline\hline\end{tabular*}
}
\renewcommand{\arraystretch}{1}
\label{tab6}
\end{table}

\begin{figure}[htbp]
\center{\includegraphics[width=0.9\linewidth]{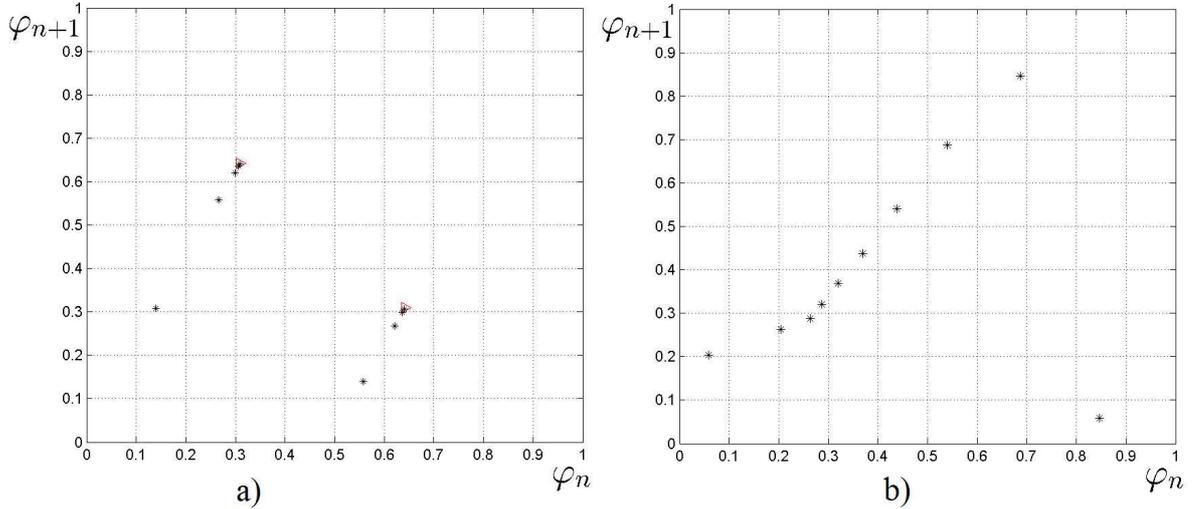}
\caption{The phase map for the cycles of the system (\ref{eq13}) at the
saddle-node bifurcation points: a) a cycle of period 9 with the code
L{\{}RL{\}}$^{4}$ at the point $A_{1}$=7.1, $h$=-1.5784680; b) a cycle of period
10 with the code RL$^{9}$ at the point $A_{1}$=3.8616612, $h$=-0.05.}
\label{Fig8}}
\end{figure}

An estimate of the phase scale factor is of the order of 1 (see Table 6),
and if we take into account that its value is directly related to the
corresponding slope of the branch of the map for the phase, then obviously
it is no longer a Bernoulli type map. This is clearly seen in Fig.~\ref{Fig8}b,
which shows a map for the phase on the cycle of period 10 at the point of
the tangent bifurcation from the Table 6. This looks more like a phase map
for a resonant cycle on a torus. For comparison, Fig.~\ref{Fig8}a shows a graph of
the mapping for the phase on the cycle of period 9 from Table 5. Here you
can see how the elements of the cycle of period 9 accumulate to the elements
of the cycle of period 2, which are indicated by triangles.

Obviously, for small $h$, the model of a one-dimensional mapping with a hole
loses its applicability here, which is connected with the fact that at the
moment of destruction of the chaotic attractor the phase dynamics is no
longer described by the Bernoulli map. This is due to the fact that a stable
cycle, which for $h$ <0 coexists with a chaotic attractor, loses stability at
$h$ = 0 via the Neimark-Sacker bifurcation and a stable torus is produced
instead. In this case, the scenario of birth/destruction of the hyperbolic
chaotic attractor also changes. For $h$> 0, it arises when a quasiperiodic
behavior is destroyed, and this transition is already not catastrophic, but
gradual (see \cite{9}). But this is a completely different story and will be
discussed elsewhere.

\bigskip

\section{Conclusions}

To describe the regularities of the chaotic set appearance under the
saddle-node scenario of the birth of the Smale--Williams hyperbolic
attractor, it is suggested using the one-dimensional Bernoulli mapping with
the ''forbidden zone''. In such mapping, a non-trivial chaotic set (with
non-zero Hausdorff dimension) arises in the general case as a result of
cascade of period-adding bifurcations characterized by geometric scaling
both in the phase space and in the parameter space.

Numerical analysis of behavior of models demonstrating the saddle-node
scenario of the birth of chaotic attractor shows that these regularities are
preserved qualitatively as we pass from 1D model to multidimensional
systems. Numerical scale exponents characterizing the average rate of
accumulation of critical cycle elements to characteristic points in the
phase space also correspond to those determined by the model. It should be
noted that these regularities are reproduced only as long as the phase
component of the dynamics continues to be described by a Bernoulli type map
at the time of the disappearance of the chaotic attractor.

The scale factor in the parameter space in all the examples considered was
significantly smaller than the values predicted by the one-dimensional
model. Apparently, this is a consequence of the assumptions made in the
derivation of the approximate mapping. Nevertheless, the nature of the
convergence of the points of bifurcations of cycles remains geometric.

\section*{Acknowledgements}

\textit{The work was supported by the grant of the Russian Scientific Foundation No 17-12-01008.
Authors acknowledge Prof. S.P.~Kuznetsov and Prof. A.~Pikovsky for useful discussion.}


\begin{thebibliography}{999}


\bibitem{1}
Smale, S. (1967), ``Differentiable dynamical systems,'' \textit{Bull. Amer. Math. Soc. }, \textbf{73},
747-817.

\bibitem{2}
Williams, R. F. (1974), ``Expanding attractors,'' \textit{Publ. Math. de l'IHES }, \textbf{43}, 169-203.

\bibitem{3}
Isaeva, O. B., Kuznetsov, S. P. and Sataev, I. R. (2012), A ''saddle-node'' bifurcation scenario for birth or destruction of a Smale--Williams solenoid, \textit{Chaos: An Interdisciplinary Journal of Nonlinear Science}, \textbf{22}(4), 043111.

\bibitem{4}
Isaeva, O. G. B., Kuznetsov, S. P., Sataev, I. R. and Pikovsky, A. S. (2013), On a bifurcation scenario of a birth of attractor of Smale--Williams type, \textit{Nelineinaya Dinamika [Russian Journal of Nonlinear Dynamics]}, \textbf{9}(2), 267-294.

\bibitem{5}
Kuznetsov, S. P. (2005), Example of a physical system with a hyperbolic attractor of the Smale--Williams type, \textit{Physical review letters}, \textbf{95}(14), 144101.

\bibitem{6}
Kuznetsov, S. P. and Sataev, I. R. (2007), Hyperbolic attractor in a system of coupled non-autonomous van der Pol oscillators: Numerical test for expanding and contracting cones, \textit{Physics Letters A}, \textbf{365}(1-2), 97-104.

\bibitem{7}
Kuznetsov, S. P. and Sataev, I. R. (2006), Verification of hyperbolicity conditions for a chaotic attractor in a system of coupled nonautonomous van der Pol oscillators, \textit{Izvestiya VUZ. Appl. Nonlin. Dynam.(Saratov)}, \textbf{14}, 3-29.

\bibitem{8}
Wilczak, D. (2010), Uniformly hyperbolic attractor of the Smale--Williams type for a Poincar{\'e} map in the Kuznetsov system, \textit{SIAM Journal on Applied Dynamical Systems}, \textbf{9}(4), 1263-1283.

\bibitem{9}
Isaeva, O. B., Kuznetsov, S. P., Sataev, I. R., Savin, D. V. and Seleznev, E. P. (2015), Hyperbolic chaos and other phenomena of complex dynamics depending on parameters in a nonautonomous system of two alternately activated oscillators, \textit{International Journal of Bifurcation and Chaos}, \textbf{25}(12), 1530033.

\bibitem{10}
Buljan, H. and Paar, V. (2001), Many-hole interactions and the average lifetimes of chaotic transients that precede controlled periodic motion, \textit{Physical Review E}, \textbf{63}(6), 066205.

\bibitem{11}
Paar, V. and Pavin, N. (1997), Missing preimages for chaotic logistic map with a hole, \textit{Fizika B}, \textbf{6}(1), 23-35.

\bibitem{12}
Paar, V. and Pavin, N. (1997), Bursts in average lifetime of transients for chaotic logistic map with a hole. \textit{Physical Review E}, \textbf{55}(4), 4112.

\bibitem{13}
Dettmann, C. (2012), Open circle maps: small hole asymptotics, \textit{Nonlinearity}, \textbf{26}(1), 307.

\bibitem{13a}
Tuval I.,Schneider J., Piro O. and Tel T. (2004), Opening up fractal structures of three-dimensional flows  via leaking, \textit{Europhysics letters}, \textbf{65}, 633. 

\bibitem{13b}
Schneider J., Tel T. and Neufeld Z. (2007), Dynamics of ``leaking'' Hamiltonian systems, \textit{Physical review E}, \textbf{66}, 066218. 

\bibitem{13c}
Altmann E.G. and Tel T. (2008), Poincar{\'e} recurrences from the perspective of transient chaos, \textit{Physical review letters}, \textbf{100}, 174101. 

\bibitem{13d}
Altmann E.G. and Tel T. (2009), Poincar{\'e} recurrences and transient chaos in systems with leaks, \textit{Physical review E}, \textbf{79}, 016204. 

\bibitem{14}
Glendinning, P. and Sidorov, N. (2015), The doubling map with asymmetrical holes, \textit{Ergodic Theory and Dynamical Systems}, \textbf{35}(4), 1208-1228.

\bibitem{15}
Sidorov, N. (2014), Supercritical holes for the doubling map, \textit{Acta Mathematica Hungarica}, \textbf{143}(2), 298-312.

\bibitem{16}
Hare, K. G. and Sidorov, N. (2014), On cycles for the doubling map which are disjoint from an interval, \textit{Monatshefte fur Mathematik}, \textbf{175}(3), 347-365.

\bibitem{17}
Procaccia, I., Thomae, S. and Tresser, C. (1987), First-return maps as a unified renormalization scheme for dynamical systems, \textit{Physical Review A}, \textbf{35}(4), 1884.
\end{thebibliography}
\end{document}